# Finite Horizon Adaptive Optimal Distributed Power Allocation for Enhanced Cognitive Radio Network in the Presence of Channel Uncertainties

Hao Xu and S. Jagannathan
Department of Electrical and Computer Engineering
Missouri University of Science and Technology
Rolla, MO, USA
hx6h7@mst.edu, sarangap@mst.edu

**ABSTRACT**

*In this paper, novel enhanced Cognitive Radio Network (CRN) is considered by using power control where secondary users (SUs) are allowed to use wireless resources of the primary users (PUs) when PUs are deactivated, but also allow SUs to coexist with PUs while PUs are activated by managing interference caused from SUs to PUs. Therefore, a novel finite horizon adaptive optimal distributed power allocation (FH-AODPA) scheme is proposed by incorporating the effect of channel uncertainties for enhanced CRN in the presence of wireless channel uncertainties under two cases. In Case 1, proposed scheme can force the Signal-to-interference (SIR) of the SUs to converge to a higher target value for increasing network throughput when PU's are not communicating within finite horizon. Once PUs activated as in the Case 2, proposed scheme can not only force the SIR's of PUs to converge to a higher target SIR, but also force the SIR's of SUs to converge to a lower value for regulating their interference to Pus during finite time period. In order to mitigate the attenuation of SIR's due to channel uncertainties the proposed novel FH-AODPA allows the SIR's of both PUs' and SUs' to converge to a desired target SIR while minimizing the energy consumption within finite horizon. Simulation results illustrate that this novel FH-AODPA scheme can converge much faster and cost less energy than others by adapting to the channel variations optimally.*

**KEYWORDS**

*Adaptive optimal distributed power allocation (AODPA), Signal-to-interference ratio (SIR), Channel uncertainties, Cognitive Radio Network*

## 1 Introduction

Cognitive Radio Network (CRN) [1] is a promising wireless network since CRN can improve the wireless resource (e.g. spectrum, power etc.) usage efficiency significantly by implementing more flexible wireless resource allocation policy [2]. In [3], a novel secondary spectrum usage scheme (i.e. opportunistic spectrum access) is introduced. The SUs in SRN can access the spectrum allocated to PUs originally while the spectrum is not used by any PU.

Moreover, the transmission power allocation plays a key role in cognitive radio network protocol designs. The efficient power allocation can not only improves the network performance (e.g. spectrum efficient, network throughput etc.), but also guarantees the Quality-of-Service (QoS) of PUs. A traditional scheme to protect transmission of PUs is introduced in [4] by imposing a power constraint less than a prescribed threshold referred to as interference temperature constraint [5] in order to contain the interference caused by SUs to each PU.

Motivated by this idea, many researchers have utilized transmission power allocation for enhanced CRN subject to interference constraint. Authors in [6] proposed a centralized power allocation (CPA) to improve the enhanced cognitive radio network performance by balancing the signal-to-interference ratios (SIR) of all PUs and SUs. However, since centralized power allocation scheme requires the information from every PU and SU which might not be possible, distributed power allocation (DPA) is preferred for enhanced CRN since information from other PUs

Research Supported by NSF ECCS#1128281 and Intelligent System Center.

and SUs is not needed. In [7-8], authors developed distributed SIR-balancing schemes to maintain Quality-of-Service requirement for each PU and SU. However, wireless channel uncertainties which are critical are not considered in these works [1-8].

For incorporating channel uncertainties into DPA, Jagannathan and Zawodniok [9] proposed a novel DPA algorithm to maintain a target SIR for each wireless receiver in cellular network under channel uncertainties. In [9], an adaptive estimator (AE) is derived to estimate slowly time varying SIR model which can be changed with varying power and channel uncertainties, and then adaptive DPA is proposed to force actual SIR of each wireless user converge to target SIR. Motivated from this idea, channel uncertainties have also been included into the developed novel finite horizon adaptive optimal DPA scheme.

In this paper, a novel finite horizon adaptive optimal distributed power allocation (FH-AODPA) for PUs and SUs in enhanced CRN with channel uncertainties is proposed. Based on the special property of enhanced CRN (i.e. introduced SUs can use PU's wireless resource when PUs are deactivated, also SUs are allowed to coexist with PUs while PUs are activated by managing interference caused from SUs to PUs properly), FH-AODPA can be developed under two cases: Case 1 PUs are deactivated while in Case 2 PUs are activated. In Case 1, since PUs are deactivated and SUs would dominant CRN, proposed FH-AODPA has to force the SIRs of SUs to converge to a higher target value in order to increase the CRN utility within finite horizon. However, in Case 2, since PUs are activated, proposed FH-AODPA has to not only force the SIRs of the SUs to converge to a low target value to guarantee QoS for PUs, but also increase network utility by allocating the transmission power properly for both PUs and SUs during finite time period.

Therefore, according to the target SIRs, the novel SIR error dynamics with channel uncertainties are derived first for PUs and SUs under two cases. Second, by using idea of adaptive dynamic programming (ADP), the novel adaptive value function estimator and finite horizon optimal DPA are proposed without known channel uncertainties for both PUs and SUs under two cases. It is important to note that proposed FH-AODPA scheme can not only forces each PU's and SU's SIR converge to target SIRs respectively in two cases, but also optimizes the power allocation during finite convergence period which is more challenging compared due to terminal state constraint. The finite horizon optimal DPA case has not been addressed so far in the literature. Compared with infinite horizon, finite horizon optimal DPA design should optimize the network utility while satisfying the terminal constraint [14]. Meanwhile, proposed FH-AODPA algorithm being highly distributive in nature does not require any inter-link communication, centralized computation, and reciprocity assumption as required in a centrally control wireless environment.

This paper is organized as follows. Section II introduces the background included cognitive radio network and wireless channel with uncertainties. Next, a novel adaptive optimal distributed power allocation (AODPA) scheme is proposed along with convergence proof for both PUs and SUs in enhanced CRN under two cases in Section III. Section IV illustrates the effectiveness of proposed adaptive optimal distributed power allocation in enhanced CRN via numerical simulations, and Section V provides concluding remarks.

## 2 Background

### 2.1 Enhanced Cognitive Radio Network (CRN)

As shown in Figure 1, the general enhanced Cognitive Radio Network (CRN) can be classified into two types of sub-networks: Primary Radio Network (PRN) and Secondary Radio Network (SRN) which include all PUs and SUs in enhanced CRN respectively. In order to improving network utilities (e.g. spectrum efficiency etc.), SUs are introduced in enhanced CRN to share the wireless resource (e.g. spectrum etc.) with PUs which usually exclusive network resource in other existing wireless networks (e.g. WLAN, WiMAX, etc.). On the other hand, similar to traditional wireless networks, QoS of PUs have to be guaranteed in enhanced CRN even though

SUs coexist. Therefore, SUs in enhanced CRN need to learn the wireless communication environment and decide their communication specifications (e.g. transmission power, target SIR, etc.) to not only maintain the QoS of PUs, but also increase the network utility such as spectrum efficiency and so on.

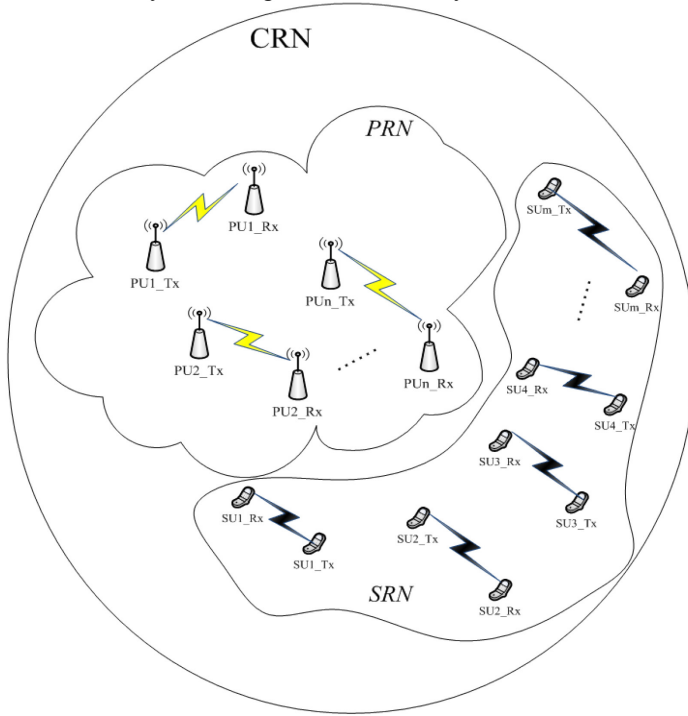

Figure 1. Enhanced Cognitive Radio Network

Due to special property of enhanced CRN, traditional wireless network protocol (e.g. resource allocation, scheduling etc.) might not be suitable for CRN. Therefore, novel protocol is extremely needed to be developed for enhanced CRN. Using the enhanced CRN property, novel protocol has to be separated into two cases: Case 1 PUs are deactivated; Case 2 PUs are activated. In Case 1, since SUs dominant the enhanced CRN, enhanced CRN network protocol has to improve the SRN performance as much as possible. However, in Case 2, enhanced CRN network protocol has to not only guarantee QoS of PUs, but also increase CRN network utility by allocating resource to both PUs and SUs properly.

**2.2 Channel uncertainties**

It is important to note that wireless channel imposes limitations on the wireless network which also includes the cognitive radio network. The wireless link between the transmitter and the receiver can be simple line-of-sight (LOS) or non LOS, or combining LOS and non LOS. In contrast to a wired channel, wireless channels are unpredictable, and harder to analyze since more uncertain elements (e.g. fading, shadowing etc.) are involved. In this paper, three main channel uncertainties (i.e. path loss, the shadowing, and Rayleigh fading) are considered. Since these main wireless channel uncertainties factors can attenuate the power of the received signal and cause variations in the SIR at the receiver significantly, it is very important to understand these before proposing novel adaptive optimal DPA scheme for cognitive radio network. The details are given as follows.

In [10], assuming path loss, received power attenuation can be expressed as the following inverse *n*-th power law

$$h_{ij} = \frac{\bar{h}}{d_{ij}^n} \qquad (1)$$

where $\bar{h}$ is a constant gain which usually is equal to 1 and $d_{ij}$ is the distance between the transmitter of $j^{th}$ user to the receiver of $i^{th}$ user and $n$ is the path loss exponent. Note that the value of path loss exponent (i.e. $n$) is depended on the characteristic of wireless communication medium. Therefore, path loss exponent $n$ have different values for different wireless propagation environments. In this paper, $n$ is set to 4 which is normally used to model path loss in the urban environment. Moreover, the channel gain $h_{ij}$ is a constant when mobility of multiple wireless users is not considered.

Further in large urban areas, high buildings, mountains and other objects can block the wireless signals and blind area can be formed behind a high rise building or between two buildings. For modeling the attenuation of the shadowing to the received power, the term $10^{0.1\zeta}$ is used to model as [11-12], where $\zeta$ is defined to be a Gaussian random variable. In next generation wireless communication system included cognitive radio network, the Rayleigh distribution is commonly used to describe the statistical time varying nature of the received envelope of a flat fading signal, or the envelope of an individual multipath component. In [10], the Rayleigh distribution has a probability density function (pdf), $p(x)$, given as:

$$p(x) = \begin{cases} \dfrac{x}{\sigma^2} \exp\left(-\dfrac{x^2}{2\sigma^2}\right) & 0 \leq x \leq \infty \\ 0 & x < 0 \end{cases} \quad (2)$$

where $x$ is a random variable, and $\sigma^2$ is known as the fading envelope of the Rayleigh distribution.

Since all of these factors (i.e. pass loss, shadowing and Rayleigh fading) can impact the power of received signals and SIR of multiple users, a channel gain factor is used and multiplied with transmitted power to present the effect from these wireless channel uncertainties. The channel gain $h$ can be derived d as [10-11]:

$$h = f(d, n, X, \zeta) = d^{-n} \cdot 10^{0.1\zeta} \cdot X^2 \quad (3)$$

where $d^{-n}$ represents the effect from path loss, $10^{0.1\zeta}$ corresponds to the effect from shadowing. For presenting Rayleigh fading, it is usually to model the power attenuation as $X^2$, where $X$ is a random variable with Rayleigh distribution. Obviously, the channel gain $h$ is a function of time.

### 3 Proposed finite horizon adaptive optimal distributed power allocation (FH-AODPA) scheme

In this section, a novel finite horizon adaptive optimal distributed power allocation (FH-AODPA) scheme is proposed to optimize the power consumption by forcing the SIRs of the PUs and SUs in enhanced CRN to converge to desired target SIRs within finite time under two cases (i.e. Case 1: PUs are deactivated, Case 2: PUs are activated) even with unknown wireless channel uncertainties respectively. In Case 1, since PUs are deactivated, proposed FH-AODPA scheme can force SUs' SIRs converge to a high target SIR in order to increasing the network capacity. In Case 2, proposed FH-AODPA scheme cannot only guarantee the PUs' communication quality by forcing PUs' SIRs converge to a desired target SIR, but also allow SUs to coexist with PUs in enhanced CRN by forcing their SIRs converge to a low target SIR which maintain the interference temperature constraints.

Next, the SIR time-varying model with unknown wireless channel uncertainties is introduced for PUs and SUs. Subsequently, we setup the value function for PUs and SUs in enhanced CRN under two cases respectively. Then, a model-free online tuning scheme is proposed to learn the value function of PUs and SUs adaptively for two cases within finite time, and then based on different cases we develop the finite horizon adaptive optimal distributed power allocation for PUs and SUs by minimizing the corresponded value function that is learned. Eventually, the convergence proof is given. Meanwhile, without loss of generality, $l\,th$ PU and $m\,th$ SU are selected to derive finite horizon adaptive optimal distributed power allocation for convenience respectively.

### 3.1 Dynamic SIR Representation for PUs and SUs with Unknown Uncertainties

In previous power allocation schemes [5-8], only path loss uncertainty is considered. In addition, without considering the mobility of PUs and SUs, the mutual interference $I(t)$ is held constant which is actually inaccurate in practical cognitive radio network. Therefore, in this paper, more uncertainties factors included path loss, shadowing and Rayleigh fading are considered together and both channel gain $h$ and the mutual interference $I(t)$ are assumed to be slowly time-varying. According to [9], the SIRs, $R_l^{PU}(t)$ $R_m^{SU}(t)$, at the receiver of $l\,th$ PU and $m\,th$ SU at the time instant $t$ can be calculated respectively as:

$$R_l^{PU}(t) = \frac{h_{ll}(t)P_l^{PU}(t)}{I_l^{PU}(t)} = \frac{h_{ll}(t)P_l^{PU}(t)}{\sum_{l\neq j\in\{PUs\}} h_{lj}(t)P_j^{PU}(t) + \sum_{i\in\{SUs\}} h_{lj}(t)P_i^{SU}(t)}$$

$$R_m^{SU}(t) = \frac{h_{mm}(t)P_m^{SU}(t)}{I_m^{SU}(t)} = \frac{h_{mm}P_m^{SU}(t)}{\sum_{j\in\{PUs\}} h_{mj}(t)P_j^{PU}(t) + \sum_{m\neq i\in\{SUs\}} h_{mi}(t)P_i^{SU}(t)}$$
(4)

where $I_l^{PU}(t), I_m^{SU}(t)$ is the mutual interference for $l\,th$ PU and $m\,th$ SU, $P_l^{PU}(t), P_m^{SU}$ are the transmitter power of $l\,th$ PU and $m\,th$ SU, and $\{PUs\}, \{SUs\}$ are the sets of PUs and SUs respectively.

Differentiating (4) on both sides, (4) can be expressed as

$$[R_l^{PU}(t)]' = \frac{dR_l^{PU}(t)}{dt} = \frac{(h_{ll}(t)P_l^{PU}(t))'I_l^{PU}(t) - (h_{ll}(t)P_l^{PU}(t))[I_l^{PU}(t)]'}{[I_l^{PU}(t)]^2}$$

$$[R_m^{SU}(t)]' = \frac{dR_m^{SU}(t)}{dt} = \frac{(h_{mm}(t)P_m^{SU}(t))'I_m^{SU}(t) - (h_{mm}(t)P_m^{SU}(t))[I_m^{SU}(t)]'}{[I_m^{SU}(t)]^2}$$
(5)

where $[R_l^{PU}(t)]', [R_m^{SU}(t)]'$ are the derivatives of $l\,th$ PU's and $m\,th$ SU's SIR (i.e. $R_l^{PU}(t), R_m^{SU}(t)$) and $[I_l^{PU}(t)]', [I_m^{SU}(t)]'$ are the derivative of $I_l^{PU}(t), I_m^{SU}(t)$. According to Euler's formula, differential equation can be transformed to discrete-time domain. Therefore, equation (5) can be expressed in discrete time by using Euler's formula as

$$(R_{l,k}^{PU})' = \frac{(h_{ll,k}P_{l,k}^{PU})'I_{l,k}^{PU} - (h_{ll,k}P_{l,k}^{PU})(I_{l,k}^{PU})'}{(I_{l,k}^{PU})^2}$$

$$(R_{m,k}^{SU})' = \frac{(h_{mm,k}P_{m,k}^{SU})'I_{m,k}^{SU} - (h_{mm,k}P_{m,k}^{SU})(I_{m,k}^{SU})'}{(I_{m,k}^{SU})^2}$$
(6)

In other words,

$$\frac{R_{l,k+1}^{PU} - R_{l,k}^{PU}}{T} = \frac{P_{l,k+1}^{PU}}{I_{l,k}^{PU}}\frac{(h_{ll,k+1} - h_{ll,k})}{T} + \frac{h_{ll,k}}{I_{l,k}^{PU}}\frac{(P_{l,k+1}^{PU} - P_{l,k}^{PU})}{T} - \frac{h_{ll,k}P_{l,k}^{PU}}{(I_{l,k}^{PU})^2}\left[\sum_{j\neq l\in\{PUs\}}\left(\frac{h_{lj,k+1} - h_{lj,k}}{T}\right.\right.$$

$$\left.\left.\times P_{j,k}^{PU} + \frac{P_{j,k+1}^{PU} - P_{j,k}^{PU}}{T}h_{lj,k}\right) + \sum_{i\in\{SUs\}}\left(\frac{h_{li,k+1} - h_{li,k}}{T}P_{l,k}^{PU} + \frac{P_{i,k+1}^{SU} - P_{i,k}^{SU}}{T}h_{lj,k}\right)\right]$$ (7)

$$\frac{R_{m,k+1}^{SU} - R_{m,k}^{SU}}{T} = \frac{P_{m,k+1}^{SU}}{I_{m,k}^{SU}}\frac{(h_{mm,k+1} - h_{mm,k})}{T} + \frac{h_{mm,k}}{I_{m,k}^{SU}}\frac{(P_{m,k+1}^{SU} - P_{m,k}^{SU})}{T} - \frac{h_{mm,k}P_{m,k}^{SU}}{I_{m,k}^2}\left[\sum_{i\neq l\in\{SUs\}}\left(\frac{h_{mi,k+1} - h_{mi,k}}{T}\right.\right.$$

$$\left.\left.\times P_{i,k}^{SU} + \frac{P_{i,k+1}^{SU} - P_{i,k}^{SU}}{T}h_{mi,k}\right) + \sum_{j\in\{PUs\}}\left(\frac{h_{mj,k+1} - h_{mj,k}}{T}P_{j,k}^{PU} + \frac{P_{j,k+1}^{PU} - P_{j,k}^{PU}}{T}h_{mj,k}\right)\right].$$

Multiplying $T$ on both sides, (7) can be derived as

$$R_{l,k+1}^{PU} = (\frac{h_{ll,k+1} - h_{ll,k}}{h_{ll,k}} - \frac{1}{I_{l,k}^{PU}}[\sum_{j \neq l \in \{PUs\}}[(h_{lj,k+1} - h_{lj,k})P_{j,k}^{PU} + (P_{j,k+1}^{PU} - P_{j,k}^{PU})h_{lj,k}]$$
$$+ \sum_{i \in \{SUs\}}[(h_{li,k+1} - h_{li,k})P_{i,k}^{SU} + (P_{i,k+1}^{SU} - P_{i,k}^{SU})h_{li,k}]])R_{l,k}^{PU} + h_{ll,k}\frac{P_{l,k+1}^{PU}}{I_{l,k}^{PU}}$$

$$R_{m,k+1}^{SU} = (\frac{h_{mm,k+1} - h_{mm,k}}{h_{mm,k}} - \frac{1}{I_{m,k}^{SU}}[\sum_{i \neq m \in \{SUs\}}[(h_{mi,k+1} - h_{mi,k})P_{i,k}^{SU} + (P_{i,k+1}^{SU} - P_{i,k}^{SU})h_{mi,k}]$$
$$+ \sum_{j \in \{PUs\}}[(h_{mj,k+1} - h_{mj,k})P_{j,k}^{PU} + (P_{j,k+1}^{PU} - P_{j,k}^{PU})h_{mj,k}]])R_{m,k}^{SU} + h_{mm,k}\frac{P_{m,k+1}^{SU}}{I_{m,k}^{SU}}$$

(8)

Next, by defining the variables $\{\phi_{l,k}^{PU}, \rho_{l,k}^{PU}, \upsilon_{l,k}^{PU}\}$ and $\{\phi_{m,k}^{SU}, \rho_{m,k}^{SU}, \upsilon_{m,k}^{SU}\}$ for $l$th PU and $m$th SU respectively as

$$\begin{cases} \phi_{l,k}^{PU} = \frac{h_{ll,k+1} - h_{ll,k}}{h_{ll,k}} - \frac{1}{I_{l,k}^{PU}}[\sum_{j \neq l \in \{PUs\}}[(h_{lj,k+1} - h_{lj,k})P_{j,k}^{PU} + (P_{j,k+1}^{PU} - P_{j,k}^{PU})h_{lj,k}] \\ \qquad + \sum_{i \in \{SUs\}}[(h_{li,k+1} - h_{li,k})P_{i,k}^{SU} + (P_{i,k+1}^{SU} - P_{i,k}^{SU})h_{li,k}]] \\ \rho_{l,k}^{PU} = h_{ll,k} \\ \upsilon_{l,k}^{PU} = \frac{P_{l,k+1}^{PU}}{I_{l,k}^{PU}} \end{cases}$$

(9)

and

$$\begin{cases} \phi_{m,k}^{SU} = \frac{h_{mm,k+1} - h_{mm,k}}{h_{mm,k}} - \frac{1}{I_{m,k}^{SU}}[\sum_{i \neq m \in \{SUs\}}[(h_{mi,k+1} - h_{mi,k})P_{i,k}^{SU} + (P_{i,k+1}^{SU} - P_{i,k}^{SU})h_{mi,k}] \\ \qquad + \sum_{j \in \{PUs\}}[(h_{mj,k+1} - h_{mj,k})P_{j,k}^{PU} + (P_{j,k+1}^{PU} - P_{j,k}^{PU})h_{mj,k}]] \\ \rho_{m,k}^{SU} = h_{mm,k} \\ \upsilon_{m,k}^{SU} = \frac{P_{m,k+1}^{SU}}{I_{m,k}^{SU}} \end{cases}$$

(10)

Then, equation (8) can be represented for $l$th PU and $m$th SU respectively as

$$R_{l,k+1}^{PU} = \phi_{l,k}^{PU} R_{l,k}^{PU} + \rho_{l,k}^{PU} \upsilon_{l,k}^{PU}$$
$$R_{m,k+1}^{SU} = \phi_{m,k}^{SU} R_{m,k}^{SU} + \rho_{m,k}^{SU} \upsilon_{m,k}^{SU}$$

(11)

Using equation (11), SIR dynamics of each PU (i.e. $l \in \{PUs\}$) and SU (i.e. $m \in \{SUs\}$) can be obtained without loss of generality. Moreover, it is observed that the SIR dynamics for PUs and SUs is a function of wireless channel variation from time instant $k$ to $k+1$. However, due to uncertainties, wireless channel variation cannot be known beforehand which causes the DPA scheme development for PUs and SUs more different and challenging, especially for finite horizon optimal designing. For solving this challenging issue, a novel finite horizon adaptive optimal DPA method is proposed as next.

### 3.2 Value function setup for finite horizon adaptive optimal DPA in enhanced CRN

As introduced above, two cases in enhanced CRN need to be considered in proposed AODPA scheme (i.e. Case 1: PUs are deactivated; Case 2: PUs are activated). The value function for PUs and SUs will be setup differently for the two cases. The details are given as follows.

*Case 1: PUs are deactivated*

In Case 1, since PUs are deactivated, wireless resource (e.g. spectrum etc.) allocated to PUs will be free. Therefore, proposed FH-AODPA scheme would force SUs' SIRs converge to a high target SIR, $\gamma_H^{SU}$, in order to improving the performance of enhanced CRN (e.g. spectrum efficiency, network capacity, etc.) within finite horizon. Therefore, the SIR error dynamics for SUs in enhanced CRN can be represented by using equation (11) as:

$$e_{m,k+1}^{SU,1} = \phi_{m,k}^{SU,1} e_{m,k}^{SU,1} + (\phi_{m,k}^{SU,1} - 1)\gamma_H^{SU} + \rho_{m,k}^{SU,1} \upsilon_{m,k}^{SU,1} \tag{12a}$$

In the other words,

$$\begin{bmatrix} e_{m,k+1}^{SU,1} \\ \gamma_H^{SU} \end{bmatrix} = \begin{bmatrix} \phi_{m,k}^{SU,1} & \phi_{m,k}^{SU,1} - 1 \\ 0 & 1 \end{bmatrix} \begin{bmatrix} e_{m,k}^{SU,1} \\ \gamma_H^{SU} \end{bmatrix} + \begin{bmatrix} \rho_{m,k}^{SU,1} \\ 0 \end{bmatrix} \upsilon_{m,k}^{SU,1} \tag{12b}$$

Moreover,

$$E_{m,k+1}^{SU,1} = A_{m,k}^{SU,1} E_{m,k}^{SU,1} + B_{m,k}^{SU,1} \upsilon_{m,k}^{SU,1} \tag{12c}$$

where SIR error in Case 1 as $e_{m,k}^{SU,1} = R_{m,k}^{SU,1} - \gamma_H^{SU}$, $\gamma_H^{SU}$ is the high target SIR for SUs under Case 1, and augmented state $E_{m,k}^{SU,1} = [e_{m,k}^{SU,1} \quad \gamma_H^{SU}]^T$. Then, according to [14] and equation (12c), the cost function for $m$ th SU in Case 1 can be defined within finite time as

$$J_{m,k}^{SU,1} = \begin{cases} (E_{m,k}^{SU,1})^T G_{m,k}^{SU,1} E_{m,k}^{SU,1} & \forall k = 0,1,...,N-1 \\ (E_{m,N}^{SU,1})^T P_{m,N}^{SU,1} E_{m,N}^{SU,1} \end{cases} \tag{13}$$

where $G_{m,k}^{SU,1} \geq 0$ is the solution of the Riccati equation [14], and $NT_s$ is the final time constraint. The optimal action dependent value function $V(\bullet)$ of $m$ th SU in Case 1 is defined as:

$$V(E_{m,k}^{SU,1}, \upsilon_{m,k}^{SU,1}) = r(E_{m,k}^{SU,1}, \upsilon_{m,k}^{SU,1}) + J_{m,k+1}^{SU,1} = [(E_{m,k}^{SU,1})^T (\upsilon_{m,k}^{SU,1})^T] \Theta_{m,k}^{SU,1} [(E_{m,k}^{SU,1})^T (\upsilon_{m,k}^{SU,1})^T]^T \tag{14}$$

with $r(E_{m,k}^{SU,1}, \upsilon_{m,k}^{SU,1}) = (E_{m,k}^{SU,1})^T Q^{SU,1} E_{m,k}^{SU,1} + (\upsilon_{m,k}^{SU,1})^T S^{SU,1} \upsilon_{m,k}^{SU,1}$, $\forall k = 0,1,...,N-1$, $Q^{SU,1}$ and $S^{SU,1}$ are positive definite matrices.

Using Bellman equation and cost function definition (13), we can formulate the following equation by substituting value-function into Bellman equation as

$$\begin{bmatrix} E_{m,k}^{SU,1} \\ \upsilon_{m,k}^{SU,1} \end{bmatrix}^T \Theta_{m,k}^{SU,1} \begin{bmatrix} E_{m,k}^{SU,1} \\ \upsilon_{m,k}^{SU,1} \end{bmatrix} = r(E_{m,k}^{SU,1}, \upsilon_{m,k}^{SU,1}) + J_{m,k+1}^{SU,1} \quad \forall k = 0,1,...,N-1 \tag{15}$$

$$= \begin{bmatrix} E_{m,k}^{SU,1} \\ \upsilon_{m,k}^{SU,1} \end{bmatrix}^T \begin{bmatrix} Q^{SU,1} + (A_{m,k}^{SU,1})^T G_{m,k+1}^{SU,1} A_{m,k}^{SU,1} & (A_{m,k}^{SU,1})^T G_{m,k+1}^{SU,1} B_{m,k}^{SU,1} \\ (B_{m,k}^{SU,1})^T G_{m,k+1}^{SU,1} A_{m,k}^{SU,1} & S^{SU,1} + (B_{m,k}^{SU,1})^T G_{m,k+1}^{SU,1} B_{m,k}^{SU,1} \end{bmatrix} \begin{bmatrix} E_{m,k}^{SU,1} \\ \upsilon_{m,k}^{SU,1} \end{bmatrix}$$

Next after incorporating the terminal constraint in the value function and (15), slowly time varying $\Theta_{m,k}^{SU,1}$ matrix can be expressed as

$$\Theta_{m,k}^{SU,1} = \begin{bmatrix} \Theta_{m,k}^{EE,SU,1} & \Theta_{m,k}^{E\upsilon,SU,1} \\ \Theta_{m,k}^{\upsilon E,SU,1} & \Theta_{m,k}^{\upsilon\upsilon,SU,1} \end{bmatrix} = \begin{bmatrix} Q^{SU,1} + (A_{m,k}^{SU,1})^T G_{m,k+1}^{SU,1} A_{m,k}^{SU,1} & (A_{m,k}^{SU,1})^T G_{m,k+1}^{SU,1} B_{m,k}^{SU,1} \\ (B_{m,k}^{SU,1})^T G_{m,k+1}^{SU,1} A_{m,k}^{SU,1} & S^{SU,1} + (B_{m,k}^{SU,1})^T G_{m,k+1}^{SU,1} B_{m,k}^{SU,1} \end{bmatrix}$$

and $\forall k = 0,1,...,N-1$

$$\Theta_{m,k}^{SU,1} = \begin{bmatrix} P_{m,N}^{SU,1} & 0 \\ 0 & 0 \end{bmatrix}.$$

Next, according to [14], the gain of the optimal power allocation for $m$ th SU under Case 1 can be represented in term of value function parameters, $\Theta_{m,k}^{SU,1}$ $\forall k = 0,1,...,N-1$, as

$$K_{m,k}^{SU,1} = [S^{SU,1} + (B_{m,k}^{SU,1})^T G_{m,k+1}^{SU,1} B_{m,k}^{SU,1}]^{-1} (B_{m,k}^{SU,1})^T G_{m,k+1}^{SU,1} A_{m,k}^{SU,1} = (\Theta_{m,k}^{\upsilon\upsilon,SU,1})^{-1} \Theta_{m,k}^{\upsilon E,SU,1} \tag{16}$$

It is important to note that even Riccati equation solution, $G_{m,k}^{SU,1}$, is known, solving the optimal design gain $K_{m,k}^{SU,1}$ for $m$ th SU under Case 1 still requires its SIR error dynamics (i.e. $A_{m,k}^{SU,1}, B_{m,k}^{SU,1}$) which cannot be known due to channel uncertainties. However, if the parameter vector $\Theta_{m,k}^{SU,1}, k = 0,1,...,N-1$ can be estimated online, then $m$ th SU's SIR error dynamic is not needed to calculate finite horizon optimal DPA gain. Meanwhile, SIR of PUs in enhanced CRN will not be considered since they are deactivated in Case 1.

*Case 2: PUs are activated*

In Case 2, proposed FH-AODPA scheme should not only force PUs' SIRs converge to a desired target SIR (i.e. $\gamma^{PU}$) for maintaining their QoS, but also force SU' SIRs converge to a low target SIR (i.e. $\gamma_L^{SU}$) in order to coexist with PUs. Therefore, the SIR error dynamics for $lth$ PU and $mth$ SU can be expressed as

$$lth\ \text{PU}: \begin{bmatrix} e_{l,k+1}^{PU,2} \\ \gamma^{PU} \end{bmatrix} = \begin{bmatrix} \phi_{l,k}^{PU,2} & \phi_{l,k}^{PU,2}-1 \\ 0 & 1 \end{bmatrix} \begin{bmatrix} e_{l,k}^{PU,2} \\ \gamma^{PU} \end{bmatrix} + \begin{bmatrix} \rho_{l,k}^{PU,2} \\ 0 \end{bmatrix} \upsilon_{l,k}^{PU,2}$$

$$mth\ \text{SU}: \begin{bmatrix} e_{m,k+1}^{SU,2} \\ \gamma_L^{SU} \end{bmatrix} = \begin{bmatrix} \phi_{m,k}^{SU,2} & \phi_{m,k}^{SU,2}-1 \\ 0 & 1 \end{bmatrix} \begin{bmatrix} e_{m,k}^{SU,2} \\ \gamma_L^{SU} \end{bmatrix} + \begin{bmatrix} \rho_{m,k}^{SU,2} \\ 0 \end{bmatrix} \upsilon_{m,k}^{SU,2}$$

(17a)

In the other words,

$$lth\ \text{PU}: \quad E_{l,k+1}^{PU,2} = A_{l,k}^{PU,2} E_{l,k}^{PU,2} + B_{l,k}^{PU,2} \upsilon_{l,k}^{PU,2}$$
$$mth\ \text{SU}: \quad E_{m,k+1}^{SU,2} = A_{m,k}^{SU,2} E_{m,k}^{SU,2} + B_{m,k}^{SU,2} \upsilon_{m,k}^{SU,2}$$

(17b)

where SIR error in Case 2 for PU and SU as $e_{l,k}^{PU,2} = R_{l,k}^{PU,2} - \gamma^{PU}$, $e_{m,k}^{SU,2} = R_{m,k}^{SU,2} - \gamma_L^{SU}$, $\gamma^{PU}, \gamma_L^{SU}$ is the desired target SIR for PUs and high target SIR for SUs under Case 2, and augmented state $E_{l,k}^{PU,2} = [e_{l,k}^{PU,2} \quad \gamma^{PU}]^T$, $E_{m,k}^{SU,2} = [e_{m,k}^{SU,2} \quad \gamma_L^{SU}]^T$. Then, according to the same theory and derivation as Case 1, slowly time varying $\Theta_{l,k}^{PU,2}, \Theta_{m,k}^{SU,2}$ matrices for PU and SU in Case 2 can be expressed respectively as:

$$\Theta_{l,k}^{PU,2} = \begin{bmatrix} \Theta_{l,k}^{EE,PU,2} & \Theta_{l,k}^{E\upsilon,PU,2} \\ \Theta_{l,k}^{\upsilon E,PU,2} & \Theta_{l,k}^{\upsilon\upsilon,PU,2} \end{bmatrix} = \begin{bmatrix} Q^{PU,2} + (A_{l,k}^{PU,2})^T G_{l,k+1}^{PU,2} A_{l,k}^{PU,2} & (A_{l,k}^{PU,2})^T G_{l,k+1}^{PU,2} B_{l,k}^{PU,2} \\ (B_{l,k}^{PU,2})^T G_{l,k+1}^{SU,2} A_{l,k}^{SU,2} & S^{PU,2} + (B_{l,k}^{PU,2})^T G_{l,k+1}^{PU,2} B_{l,k}^{PU,2} \end{bmatrix}$$

$$\Theta_{m,k}^{SU,2} = \begin{bmatrix} \Theta_{m,k}^{EE,SU,2} & \Theta_{m,k}^{E\upsilon,SU,2} \\ \Theta_{m,k}^{\upsilon E,SU,2} & \Theta_{m,k}^{\upsilon\upsilon,SU,2} \end{bmatrix} = \begin{bmatrix} Q^{SU,2} + (A_{m,k}^{SU,2})^T G_{m,k+1}^{SU,2} A_{m,k}^{SU,2} & (A_{m,k}^{SU,2})^T G_{m,k+1}^{SU,2} B_{m,k}^{SU,2} \\ (B_{m,k}^{SU,2})^T G_{m,k+1}^{SU,2} A_{m,k}^{SU,2} & S^{SU,1} + (B_{m,k}^{SU,2})^T G_{m,k+1}^{SU,2} B_{m,k}^{SU,2} \end{bmatrix}$$

and $\forall k = 0,1,...,N-1$

$$\Theta_{l,N}^{PU,2} = \begin{bmatrix} P_{l,N}^{PU,2} & 0 \\ 0 & 0 \end{bmatrix}, \Theta_{m,N}^{SU,2} = \begin{bmatrix} P_{m,N}^{SU,2} & 0 \\ 0 & 0 \end{bmatrix}.$$

Next, the gain of the optimal power allocation for $lth$ PU and $mth$ SU under Case 2 can be represented in term of value function parameters, $\Theta_{l,k}^{PU,2}$, $\Theta_{m,k}^{SU,2}$, $\forall k = 0,1,...,N-1$, respectively as

$$K_{l,k}^{PU,2} = [S^{PU,2} + (B_{l,k}^{PU,2})^T G_{l,k+1}^{PU,2} B_{l,k}^{PU,2}]^{-1} (B_{l,k}^{PU,2})^T G_{l,k+1}^{PU,2} A_{l,k}^{PU,2} = (\Theta_{l,k}^{\upsilon\upsilon,PU,2})^{-1} \Theta_{l,k}^{\upsilon E,PU,2}$$
$$K_{m,k}^{SU,2} = [S^{SU,2} + (B_{m,k}^{SU,2})^T G_{m,k+1}^{SU,2} B_{m,k}^{SU,2}]^{-1} (B_{m,k}^{SU,2})^T G_{m,k+1}^{SU,2} A_{m,k}^{SU,2} = (\Theta_{m,k}^{\upsilon\upsilon,SU,2})^{-1} \Theta_{m,k}^{\upsilon E,SU,2}$$

(18)

Similar to Case 1, once value function parameters $\Theta_{l,k}^{PU,2}$, $\Theta_{m,k}^{SU,2}$ for $lth$ PU and $mth$ SU under Case 2 have been tuned, the finite horizon optimal power allocation can be obtained for PU and SU in Case 2 by using (18).

### 3.3 Model-free online tuning adaptive estimator for value function

In this section, proposed finite horizon adaptive optimal approach derives a novel estimator which is used to estimate the value function for PUs and SUs in CRN under two cases respectively. After that, finite horizon optimal power allocation design will be derived by using learned value function parameters for PUs and SUs under two cases. The details are given as follows.

#### 3.3.1 Adaptive estimator of value function and $\Theta$ matrix under two cases

*Case 1: PUs are deactivated*

Before deriving the adaptive estimator (AE), the following assumption is asserted.

**Assumption 1 [15]:** The value function, $V(E_{m,k}^{SU,1}, \upsilon_{m,k}^{SU,1})$, can be represented as the linear in the unknown parameter (LIP).

Then, using adaptive control theory [19] and (14), the value-function for *mth* SU in enhanced CRN under Case 1 can be represented in vector form as

$$V(E_{m,k}^{SU,1}, \upsilon_{m,k}^{SU,1}) = (z_{m,k}^{SU,1})^T \Theta_{m,k}^{SU,1} z_{m,k}^{SU,1} = (\theta_{m,k}^{SU,1})^T \bar{z}_{m,k}^{SU,1} = (W_m^{SU,1})^T \sigma(N-k) \bar{z}_{m,k}^{SU,1} \quad \forall k = 0,...,N \quad (19)$$

where $\theta_{m,k}^{SU,1} = vec(\Theta_{m,k}^{SU,1}) = (W_m^{SU,1})^T \sigma(N-k), z_{m,k}^{SU,1} = [(E_{m,k}^{SU,1})^T \quad \upsilon_m^T(E_{m,k}^{SU,1})]^T$ and $\bar{z}_{m,k}^{SU,1} = [(z_{m,k1}^{SU,1})^2,...,z_{m,kn-1}^{SU,1} z_{m,kn}^{SU,1}, (z_{m,kn}^{SU,1})^2]$ is the Kronecker product quadratic polynomial basis vector [20] for *mth* SU in enhanced CRN under Case 1, $vec(\bullet)$ function is constructed by stacking the columns of matrix into one column vector with off-diagonal elements [16]. Moreover, $\sigma(\bullet)$ is the time-dependent regression function for the value function parameter estimation $\theta_{m,k}^{SU,1}$. It is important to note that $\theta_{m,N}^{SU,1} = vec(\Theta_{m,N}^{SU,1})$ is considered as the known terminal constraint in finite horizon optimal DPA problem. Therefore, it is obvious that target parameter $W_m^{SU,1}$ for *mth* SU and regression function $\sigma(\bullet)$ should satisfy $\theta_{m,N}^{SU,1} = (W_m^{SU,1})^T \sigma(0)$.

Based on relationship between value function and cost function [16], cost function of *mth* SU under Case 1 can also be represented in term of $\Theta_{m,k}^{SU,1}$ as

$$\begin{aligned} J_{m,k}^{SU,1}(E) &= V(E_{m,k}^{SU,1}, \upsilon_{m,k}^{SU,1}) = (z_{m,k}^{SU,1})^T \Theta_{m,k}^{SU,1} z_{m,k}^{SU,1} = (\theta_{m,k}^{SU,1})^T \bar{z}_{m,k}^{SU,1} \\ &= (W_m^{SU,1})^T \sigma(N-k) \bar{z}_{m,k}^{SU,1} \quad \forall k = 0,...,N \end{aligned} \quad (20)$$

Next, the value function of *mth* SU in Case 1, $V(E_{m,k}^{SU,1}, \upsilon_{m,k}^{SU,1})$, can be approximated by using adaptive estimator in terms of estimated parameter $\hat{\Theta}_{m,k}^{SU,1}$ as

$$\hat{V}(E_{m,k}^{SU,1}, \upsilon_{m,k}^{SU,1}) = (\hat{\theta}_{m,k}^{SU,1})^T \bar{z}_{m,k}^{SU,1} = (\hat{W}_{m,k}^{SU,1})^T \sigma(N-k) \bar{z}_{m,k}^{SU,1} \quad \forall k = 0,...,N \quad (21)$$

where $\hat{W}_{m,k}^{SU,1}$ is the estimated value of *mth* SU's target parameter vector $\theta_{m,k}^{SU,1}$ at time $kT_s$ under Case 1.

It is observed that *mth* SU's Bellman equation in Case 1 can be rewritten as $J_{m,k+1}^{SU,1}(E) - J_{m,k}^{SU,1}(E) + r(E_{m,k}^{SU,1}, \upsilon_{m,k}^{SU,1}) = 0$. However, this relationship does not hold when we apply the estimated matrix $\hat{\Theta}_{m,k}^{SU,1}$. Hence, using delayed values for convenience; the residual error associate with (21) can be expressed as $\hat{J}_{m,k}^{SU,1}(E) - \hat{J}_{m,k-1}^{SU,1}(E) + r(E_{m,k-1}^{SU,1}, \upsilon_{m,k-1}^{SU,1}) = e_{m,Wk}^{FTBE}$, i.e.

$$\begin{aligned} e_{m,Wk}^{FTBE} &= \hat{J}_{m,k}^{SU,1}(E) - \hat{J}_{m,k-1}^{SU,1}(E) + r(E_{m,k-1}^{SU,1}, \upsilon_{m,k-1}^{SU,1}) \\ &= r(E_{m,k-1}^{SU,1}, \upsilon_{m,k-1}^{SU,1}) + (\hat{W}_{m,k}^{SU,1})^T \Delta Z_{m,k-1}^{SU,1} \quad k = 0,1,...,N \end{aligned} \quad (22a)$$

where $\Delta Z_{m,k-1}^{SU,1} = \sigma(N-k) \bar{z}_{m,k}^{SU,1} - \sigma(N-k+1) \bar{z}_{m,k-1}^{SU,1}$, and $e_{m,Wk}^{FTBE}$ is Bellman equation residual error for the finite horizon scenario of *mth* SU under Case 1. Next, the dynamics of (22a) can be represented as

$$e_{m,Wk+1}^{FTBE} = r(E_{m,k}^{SU,1}, \upsilon_{m,k}^{SU,1}) + (\hat{W}_{m,k+1}^{SU,1})^T \Delta Z_{m,k}^{SU,1} \quad \forall k = 0,1,...,N \quad (22b)$$

Besides considering $e_{m,Wk}^{FTBE}$, the estimation error $e_{m,Wk}^{FC}$ due to the terminal constraint needs to be considered and therefore given

$$e_{m,Wk}^{FC} = \theta_{m,N}^{SU,1} - (\hat{W}_{m,k}^{SU,1})^T \sigma(0) \quad (23)$$

Next, we define the auxiliary residual error vector and terminal constraint estimation error vector can be defined as

$$\begin{aligned} \Xi_{m,k}^{FTBE,SU,1} &= \Gamma_{m,k-1}^{SU,1} + (\hat{W}_{m,k}^{SU,1})^T \Delta \mathbf{Z}_{m,k-1}^{SU,1} \\ \Xi_{m,k}^{FC,SU,1} &= \theta_{m,N}^{SU,1} \Lambda - (\hat{W}_{m,k}^{SU,1})^T \sigma(0) \Lambda \end{aligned} \quad (24)$$

where $\Lambda$ is a known dimension matching matrix $\Gamma_{m,k-1}^{SU,1} = [r(E_{m,k-1}^{SU,1}, \upsilon_{m,k-1}^{SU,1}) .... r(E_{m,k-1-i}^{SU,1}, \upsilon_{m,k-1-i}^{SU,1})]$ and $\Delta \mathbf{Z}_{m,k-1}^{SU,1} = [\Delta Z_{m,k-1}^{SU,1} .... \Delta Z_{m,k-1-i}^{SU,1}]$, $0 < i < k-1$ and $\forall m \in \{SUs\}$. Then the dynamics of auxiliary residual error vector (24) are generated similar as (23) and revealed to be $\Xi_{m,k+1}^{FTBE,SU,1} = \Gamma_{m,k}^{SU,1} + (\hat{\theta}_{m,k+1}^{SU,1})^T \Delta \mathbf{Z}_{m,k}^{SU,1}$, $\Xi_{m,k+1}^{FC,SU,1} = \theta_{m,N}^{SU,1} \Lambda - (\hat{W}_{m,k+1}^{SU,1})^T \sigma(0) \Lambda$.

To force both the Bellman equation and terminal constraint estimation error converge to zero, the update law of the *mth* SU's time varying matrix $\hat{\Theta}_{m,k}^{SU,1}$ in Case 1 can be derived as

$$\hat{W}_{m,k+1}^{SU,1} = \Psi_{m,k}^{SU,1} [(\Psi_{m,k}^{SU,1})^T \Psi_{m,k}^{SU,1}]^{-1} [\alpha_W^{SU,1} (\Xi_{m,k}^{FTBE,SU,1})^T + \alpha_W^{SU,1} (\Xi_{m,k}^{FC,SU,1})^T - (\Gamma_{m,k}^{SU,1})^T - \theta_{m,N}^{SU,1} \Lambda] \quad (25)$$

where $\Psi_{m,k}^{SU,1} = \Delta \mathbf{Z}_{m,k}^{SU,1} - \sigma(0) \Lambda$ and $0 < \alpha_W^{SU,1} < 1$. Substituting (25) into (24) results

$$\Xi_{m,k+1}^{FTBE,SU,1} + \Xi_{m,k+1}^{FC,SU,1} = \alpha_W^{SU,1} (\Xi_{m,k}^{FTBE,SU,1} + \Xi_{m,k}^{FC,SU,1}) \quad (26)$$

Then defining the parameter estimation error of *mth* SU under Case 1 as $\widetilde{W}_{m,k}^{SU,1} = W_m^{SU,1} - \hat{W}_{m,k}^{SU,1}$, dynamics of estimation errors of *mth* SU's adaptive estimator parameter in Case 1 can be expressed as

$$(\widetilde{W}_{m,k+1}^{SU,1})^T [\Delta \mathbf{Z}_{m,k}^{SU,1} - \sigma(0)] = \alpha_W^{SU,1} (\widetilde{W}_{m,k}^{SU,1})^T [\Delta \mathbf{Z}_{m,k-1}^{SU,1} - \sigma(0)] \quad (27)$$

Next, the estimation of *mth* SU's optimal design under Case 1 will be derived based on tuned parameter $\hat{\Theta}_{m,k}^{SU,1}$ as

$$\hat{\upsilon}_{m,k}^{SU,1} = -\hat{K}_{m,k}^{SU,1} z_{m,k}^{SU,1} = -(\hat{\Theta}_{m,k}^{\upsilon\upsilon,SU,1})^{-1} \hat{\Theta}_{m,k}^{\upsilon E,SU,1} z_{m,k}^{SU,1} \quad (28a)$$

Then, using (10), the *mth* SU's adaptive optimal DPA design under Case 1 can be expressed as

$$P_{m,k+1}^{SU,1} = \hat{\upsilon}_{m,k}^{SU,1} I_{m,k}^{SU,1} = -(\hat{\Theta}_{m,k}^{\upsilon\upsilon,SU,1})^{-1} \hat{\Theta}_{m,k}^{\upsilon E,SU,1} z_{m,k}^{SU,1} I_{m,k}^{SU,1} \quad (28b)$$

***Case 2: PUs are activated***

Similar to Case 1, the value-function for *mth* PU and *mth* SU in CRN under Case 2 can be estimated in vector form respectively as

$$\begin{aligned} lth \text{ PU}: \hat{V}(E_{l,k}^{PU,2}, \upsilon_{l,k}^{PU,2}) &= (\hat{\theta}_{l,k}^{PU,2})^T \bar{z}_{l,k}^{PU,2} = (\hat{W}_{l,k}^{PU,2})^T \sigma(N-k) \bar{z}_{l,k}^{PU,2} \\ mth \text{ SU}: \hat{V}(E_{m,k}^{SU,2}, \upsilon_{m,k}^{SU,2}) &= (\hat{\theta}_{m,k}^{SU,2})^T \bar{z}_{m,k}^{SU,2} = (\hat{W}_{m,k}^{SU,2})^T \sigma(N-k) \bar{z}_{m,k}^{SU,2} \end{aligned} \quad (29)$$

where $\bar{z}_{l,k}^{PU,2}, \bar{z}_{m,k}^{SU,2}$ are Kronecker product quadratic polynomial basis vector for *lth* PU and *mth* SU in enhanced CRN under Case 2.

Next, the update law of *lth* PU's and *mth* SU's time varying matrices, $\hat{\Theta}_{l,k}^{PU,2}, \hat{\Theta}_{m,k}^{SU,2}$, in Case 2 can be derived respectively as

$$\begin{aligned} lth \text{ PU}: \hat{W}_{l,k+1}^{PU,2} &= \Psi_{l,k}^{PU,2} [(\Psi_{l,k}^{PU,2})^T \Psi_{l,k}^{PU,2}]^{-1} [\alpha_W^{PU,2} (\Xi_{l,k}^{FTBE,PU,2})^T + \alpha_W^{PU,2} (\Xi_{l,k}^{FC,PU,2})^T \\ &\quad - (\Gamma_{l,k}^{PU,2})^T - \theta_{m,N}^{PU,2} \Lambda] \\ mth \text{ SU}: \hat{W}_{m,k+1}^{SU,2} &= \Psi_{m,k}^{SU,2} [(\Psi_{m,k}^{SU,2})^T \Psi_{m,k}^{SU,2}]^{-1} [\alpha_W^{SU,2} (\Xi_{m,k}^{FTBE,SU,2})^T + \alpha_W^{SU,2} (\Xi_{m,k}^{FC,SU,2})^T \\ &\quad - (\Gamma_{m,k}^{SU,2})^T - \theta_{m,N}^{SU,2} \Lambda] \end{aligned} \quad (30)$$

where tuning parameter $0 < \alpha_W^{PU,2} < 1$ and $0 < \alpha_W^{SU,2} < 1$. Then, dynamics of estimation errors for *lth* PU's and *mth* SU's adaptive estimator parameter in Case 2 can be represented respectively as

$$\begin{aligned} lth \text{ PU}: &\quad (\widetilde{W}_{l,k+1}^{PU,2})^T \Delta \mathbf{Z}_{l,k}^{PU,2} = \alpha_\theta^{PU,2} (\widetilde{W}_{l,k}^{PU,2})^T \Delta \mathbf{Z}_{l,k}^{PU,2} \\ mth \text{ SU}: &\quad (\widetilde{W}_{m,k+1}^{SU,2})^T \Delta \mathbf{Z}_{m,k}^{SU,2} = \alpha_\theta^{SU,2} (\widetilde{W}_{m,k}^{SU,2})^T \Delta \mathbf{Z}_{m,k}^{SU,2} \end{aligned} \quad (31)$$

Then, we can derive the finite horizon adaptive optimal DPA design for *lth* PU and *mth* SU in enhanced CRN under Case 2 based on tuned parameter $\hat{\Theta}_{l,k}^{PU,2}, \hat{\Theta}_{m,k}^{SU,2}$, $\forall k = 0,1,..., N-1$, respectively as:

$$\begin{aligned} lth \text{ PU}: &\quad P_{l,k+1}^{PU,2} = \hat{\upsilon}_{l,k}^{PU,2} I_{l,k}^{PU,2} = -(\hat{\Theta}_{l,k}^{\upsilon\upsilon,PU,2})^{-1} \hat{\Theta}_{l,k}^{\upsilon E,PU,2} z_{l,k}^{PU,2} I_{l,k}^{PU,2} \\ mth \text{ SU}: &\quad P_{m,k+1}^{SU,2} = \hat{\upsilon}_{m,k}^{SU,2} I_{m,k}^{SU,2} = -(\hat{\Theta}_{m,k}^{\upsilon\upsilon,SU,2})^{-1} \hat{\Theta}_{m,k}^{\upsilon E,SU,2} z_{m,k}^{SU,2} I_{m,k}^{SU,2} \end{aligned} \quad (32)$$

Eventually, the stability of value function estimation, adaptive DPA estimation, and adaptive estimation error dynamics for PUs and SUs in two cases are considered in next section.

**3.3.2 Closed-loop finite horizon adaptive optimal DPA system stability for PUs and SUs in enhanced CRN**

Since proposed finite horizon adaptive optimal DPA is designed for enhanced CRN PUs and SUs in two cases, the closed-loop stability will be analyzed under two cases respectively.

*Case 1: PUs are deactivated*

In this case, it will be shown that $mth$ SU's time-varying matrix, $\Theta_{l,k}$, $\forall k = 0,1,...,N-1$, and related value function estimation errors dynamic are Uniformly Ultimately Boundednes (*UUB*) when PUs in enhanced CRN are deactivated. Further, the estimated finite horizon adaptive optimal distributed power allocation will approach the optimal power allocation within a small ultimate bound. Next the initial system states (i.e. SIR errors of SUs) are considered to reside in the compact set which in turn is stabilized by using the initial stabilizing input $\upsilon_{0m,k}^{SU,1}$. Further sufficient condition for the adaptive estimator tuning gain $\alpha_W^{SU,1}$ is derived to ensure the all future SUs' SIR errors will converge close to zero. Then it can be shown that the actual finite horizon adaptive DPA approaches the optimal power allocation for SUs in Case 1 within ultimate bound during finite time period.

Before introducing the convergence proof, the algorithm represented the proposed finite horizon adaptive optimal distributed power allocation is given as follows.

---

**Algorithm 1:** Finite Horizon Adaptive Optimal Distributed Power Allocation for $mth$ SU in enhanced CRN under Case 1 (i.e. PUs are deactivated)

---

1: **Initialize:** $\hat{W}_{m,k}^{SU,1} = \mathbf{0}$ and implementing admissible policy $\upsilon_{m,0}^{SU,1}$.

2: **while** $\{ kT_s < t < (k+1)T_s \}$ **do**

3:     **Calculate** the value function estimation errors $\Xi_{m,k}^{FTBE,SU,1}$ and $\Xi_{m,k}^{FC,SU,1}$.

4:     **Update** the parameters of the value function estimator

5:     $\hat{W}_{m,k+1}^{SU,1} = \Psi_{m,k}^{SU,1}[(\Psi_{m,k}^{SU,1})^T \Psi_{m,k}^{SU,1}]^{-1}[\alpha_W^{SU,1}(\Xi_{m,k}^{FTBE,SU,1})^T + \alpha_W^{SU,1}(\Xi_{m,k}^{FC,SU,1})^T - (\Gamma_{m,k}^{SU,1})^T - \theta_{m,N}^{SU,1}\Lambda]$

6:     **Update** finite horizon adaptive optimal DPA based on estimated $\Theta_{m,k}^{SU,1}$ matrix.

7:     $\hat{\upsilon}_{m,k}^{SU,1} = -\hat{K}_{m,k}^{SU,1} z_{m,k}^{SU,1} = -(\hat{\Theta}_{m,k}^{\upsilon\upsilon,SU,1})^{-1}\hat{\Theta}_{m,k}^{\upsilon E,SU,1} z_{m,k}^{SU,1}$

8:     $P_{m,k+1}^{SU,1} = \hat{\upsilon}_{m,k}^{SU,1} I_{m,k}^{SU,1} = -(\hat{\Theta}_{m,k}^{\upsilon\upsilon,SU,1})^{-1}\hat{\Theta}_{m,k}^{\upsilon E,SU,1} z_{m,k}^{SU,1} I_{m,k}^{SU,1}$

9: **end while**

10: **If** $\{t < NT_s\}$ **do**

11:     **Go to** next time interval $[(k+1)T_s, (k+2)T_s)$ (i.e. $k = k+1$), and then go back

12:     to line 2.

13: **else do**

14:     **Stop** the algorithm.

---

**Theorem 1.** *(Convergence of the Adaptive Optimal Distributed Power Allocation for SUs in enhanced CRN under Case 1).* Let $\upsilon_{m,0}^{SU,1}$ be any initial admissible policy for the $mth$ SU's finite horizon adaptive optimal DPA scheme in Case 1 with $0 < l_o < 1/2$. Within the time horizon (i.e. $t \in [0, NT_s]$), let the $mth$ SU's parameters be tuned and estimation finite horizon optimal power allocation be provided by (25) and (28b) respectively. Then, there exists positive constant $\alpha_W^{SU,1}$ given as $0 < \alpha_W^{SU,1} < 1$ such that the $mth$ SU's SIR error $e_{m,k}^{SU,1}$ and value function parameter estimation errors $\tilde{W}_{m,k}^{SU,1}$ are all uniformly ultimately bounded (*UUB*) in Case 1 (i.e. PUs are deactivated) within the finite time horizon. Moreover, the ultimate bounds are depend on final time (i.e.

$NT_s$), bounded initial value function estimation error $B_{m,0}^{V,SU,1}$ and bounded initial SIR error state $B_{m,0}^{E,SU,1}$.

***Proof:*** Consider the following positive definite Lyapunov function candidate

$$L_k = L_D\left(E_{m,k}^{SU,1}\right) + L_J\left(\widetilde{W}_{m,k}^{SU,1}\right) \tag{33}$$

where $L_D\left(E_{m,k}^{SU,1}\right)$ is defined as $L_D\left(E_{m,k}^{SU,1}\right) = (E_{m,k}^{SU,1})^T \Pi E_{m,k}^{SU,1}$ with $\Pi = \dfrac{S^{SU,1}[1-(\alpha_W^{SU,1})^2]}{2(B_{m,M}^{SU,1})^2}\mathbf{I}$ is positive definite matrix and $\mathbf{I}$ is identity matrix, $\left\|B_{m,k}^{SU,1}\right\| \leq B_{m,M}^{SU,1}$, and $L_J(\widetilde{W}_{m,k}^{SU,1})$ is defined as

$$\begin{aligned}L_J(\widetilde{W}_{m,k}^{SU,1}) &= [(\widetilde{W}_{m,k}^{SU,1})^T \sigma(N-k)\bar{z}_{m,k}^{SU,1} - (\widetilde{W}_{m,k}^{SU,1})^T \sigma(N-k+1)\bar{z}_{m,k-1}^{SU,1}]^2 \\ &= [(\widetilde{W}_{m,k}^{SU,1})^T \Delta Z_{m,k-1}^{SU,1}]^2\end{aligned} \tag{34}$$

The first difference of (34) can be expressed as $\Delta L = \Delta L_D\left(E_{m,k}^{SU,1}\right) + \Delta L_J(\widetilde{W}_{m,k}^{SU,1})$, and considering that $\Delta L_J\left(\widetilde{W}_{m,k}^{SU,1}\right) = [(\widetilde{W}_{m,k+1}^{SU,1})^T \Delta Z_{l,k}]^2 - [(\widetilde{W}_{m,k}^{SU,1})^T \Delta Z_{l,k-1}]^2$ with value function estimator, we have

$$\begin{aligned}\Delta L_J(\widetilde{W}_{m,k}^{SU,1}) &= [(\widetilde{W}_{m,k+1}^{SU,1})^T \Delta Z_{l,k}]^2 - [(\widetilde{W}_{m,k}^{SU,1})^T \Delta Z_{l,k-1}]^2 \\ &= -[1-(\alpha_W^{SU,1})^2][(\widetilde{W}_{m,k}^{SU,1})^T \Delta Z_{m,k-1}^{SU,1}]^2 \\ &\leq -[1-(\alpha_W^{SU,1})^2]\left\|\Delta Z_{m,k-1}^{SU,1}\right\|^2 \left\|\widetilde{W}_{m,k}^{SU,1}\right\|^2\end{aligned} \tag{35}$$

Next considering the first part of Lyapunov candidate function $\Delta L_D(E_{m,k}^{SU,1}) = (E_{m,k+1}^{SU,1})^T \Pi E_{m,k+1}^{SU,1} - (E_{m,k}^{SU,1})^T \Pi E_{m,k}^{SU,1}$ and applying the proposed FH-AODPA scheme and Cauchy-Schwartz inequality reveals

$$\begin{aligned}\Delta L_D(E_{m,k}^{SU,1}) &\leq \|\Pi\|\left\|A_{m,k}^{SU,1} E_{m,k}^{SU,1} + B_{m,k}^{SU,1}\hat{\upsilon}_{m,k}^{SU,1} - B_{m,k}^{SU,1}\widetilde{\upsilon}_{m,k}^{SU,1}\right\|^2 - (E_{m,k}^{SU,1})^T \Pi E_{m,k}^{SU,1} \\ &\leq 2\|\Pi\|\left\|A_{m,k}^{SU,1} E_{m,k}^{SU,1} + B_{m,k}^{SU,1}\hat{\upsilon}_{m,k}^{SU,1}\right\|^2 + 2\|\Pi\|\left\|B_{m,k}^{SU,1}\widetilde{\upsilon}_{m,k}^{SU,1}\right\|^2 - (E_{m,k}^{SU,1})^T \Pi E_{m,k}^{SU,1} \\ &\leq -(1-2l_o)\|\Pi\|\left\|E_{m,k}^{SU,1}\right\|^2 + 2\|\Pi\|\left\|B_{m,k}^{SU,1}\widetilde{\upsilon}_{m,k}^{SU,1}\right\|^2\end{aligned} \tag{36}$$

At final step, combining the equation (34) and (36), we have

$$\begin{aligned}\Delta L_k &\leq -(1-2k^*)\|\Pi\|\left\|E_{m,k}^{SU,1}\right\|^2 + 2\|\Pi\|\left\|B_{m,k}^{SU,1}\widetilde{\upsilon}_{m,k}^{SU,1}\right\|^2 + (\alpha_W^{SU,1})^2\left\|\Delta Z_{m,k-1}^{SU,1}\right\|^2\left\|\widetilde{W}_{m,k}^{SU,1}\right\|^2 \\ &\quad - \left\|\Delta Z_{m,k-1}^{SU,1}\right\|^2\left\|\widetilde{W}_{m,k}^{SU,1}\right\|^2 \\ &\leq -(1-2k^*)\|\Pi\|\left\|E_{m,k}^{SU,1}\right\|^2 + \frac{1}{2}[1-(\alpha_W^{SU,1})^2]\left\|\Delta Z_{m,k-1}^{SU,1}\right\|^2\left\|\widetilde{W}_{m,k}^{SU,1}\right\|^2 \\ &\quad + (\alpha_W^{SU,1})^2\left\|\Delta Z_{m,k-1}^{SU,1}\right\|^2\left\|\widetilde{W}_{m,k}^{SU,1}\right\|^2 - \left\|\Delta Z_{m,k-1}^{SU,1}\right\|^2\left\|\widetilde{W}_{m,k}^{SU,1}\right\|^2 \\ &\leq -(1-2k^*)\|\Pi\|\left\|E_{m,k}^{SU,1}\right\|^2 - \frac{1}{2}[1-(\alpha_W^{SU,1})^2]\left\|\Delta Z_{m,k-1}^{SU,1}\right\|^2\left\|\widetilde{W}_{m,k}^{SU,1}\right\|^2\end{aligned} \tag{37}$$

Since $0 < k^* < 1/2$ and $0 < \alpha_W^{SU,1} < 1$, $\Delta L_k$ is negative definite and $L_k$ is positive definite. Using standard Lyapunov theory [15], during finite horizon, all the signals can be proven *UUB* with ultimate bounds are dependent on initial conditions and final time $NT_s$. The details are demonstrated as following.

Assume SIR error state and value function estimation error are initiated as a bound $B_{m,0}^{E,SU,1}$, $B_{m,0}^{V,SU,1}$ respectively (i.e. $\left\|E_{m,0}^{SU,1}\right\|^2 = B_{m,0}^{E,SU,1}$, $\left\|\Delta Z_{m,k-1}^{SU,1}\right\|^2\left\|\widetilde{W}_{m,k}^{SU,1}\right\|^2 = B_{m,0}^{E,SU,1}$). Using standard Lyapunov Theory [15], $\left\|E_{m,0}^{SU,1}\right\|$ and $\left\|\Delta Z_{m,k-1}^{SU,1}\right\|^2\left\|\widetilde{W}_{m,k}^{SU,1}\right\|^2$ for $k = 1,2,...,N$ can be represented as

$$\|\Pi\| \|E_{m,k}^{SU,1}\|^2 + \|\Delta Z_{m,k-1}^{SU,1}\|^2 \|\widetilde{W}_{m,k}^{SU,1}\|^2 = \|\Pi\| \|E_{m,0}^{SU,1}\|^2 + \|\Delta Z_{m,-1}^{SU,1}\|^2 \|\widetilde{W}_{m,0}^{SU,1}\|^2$$

$$+ \|\Pi\| \underbrace{\left(\|E_{m,1}^{SU,1}\|^2 - \|E_{m,0}^{SU,1}\|^2\right)}_{\Delta L_D(E_{m,0}^{SU,1})} + \underbrace{\left(\|\Delta Z_{m,0}^{SU,1}\|^2 \|\widetilde{W}_{m,1}^{SU,1}\|^2 - \|\Delta Z_{m,-1}^{SU,1}\|^2 \|\widetilde{W}_{m,0}^{SU,1}\|^2\right)}_{\Delta L_J(\widetilde{W}_{m,0}^{SU,1})} + \cdots \quad (38)$$

$$+ \|\Pi\| \underbrace{\left(\|E_{m,k}^{SU,1}\|^2 - \|E_{m,k-1}^{SU,1}\|^2\right)}_{\Delta L_D(E_{m,k-1}^{SU,1})} + \underbrace{\left(\|\Delta Z_{m,k-1}^{SU,1}\|^2 \|\widetilde{W}_{m,k}^{SU,1}\|^2 - \|\Delta Z_{m,k-2}^{SU,1}\|^2 \|\widetilde{W}_{m,k-1}^{SU,1}\|^2\right)}_{\Delta L_J(\widetilde{W}_{m,k-1}^{SU,1})}$$

$$\leq \|\Pi\| (B_{m,0}^{E,SU,1})^2 + (B_{m,0}^{V,SU,1})^2 + \underbrace{\Delta L_D(E_{m,0}^{SU,1}) + \Delta L_J(\widetilde{W}_{m,0}^{SU,1})}_{\Delta L_0} + \cdots + \underbrace{\Delta L_D(E_{m,k-1}^{SU,1}) + \Delta L_J(\widetilde{W}_{m,k-1}^{SU,1})}_{\Delta L_{k-1}}$$

$$\leq \|\Pi\| (B_{m,0}^{E,SU,1})^2 + (B_{m,0}^{V,SU,1})^2 + \sum_{i=0}^{k-1} \Delta L_i \qquad \forall k = 1,2,...,N$$

Using (37) and property of geometric sequence [17], equation (38) can be represented as

$$\|\Pi\| \|E_{m,k}^{SU,1}\|^2 + \|\Delta Z_{m,k-1}^{SU,1}\|^2 \|\widetilde{W}_{m,k}^{SU,1}\|^2 \leq \|\Pi\| (B_{m,0}^{E,SU,1})^2 + (B_{m,0}^{V,SU,1})^2 + \sum_{i=0}^{k-1} \Delta L_i$$

$$\leq \|\Pi\| (B_{m,0}^{E,SU,1})^2 + (B_{m,0}^{V,SU,1})^2 - \sum_{i=0}^{k-1} (1-2l_o) \|\Pi\| \|E_{m,k}^{SU,1}\|^2 - \frac{1}{2} \sum_{i=0}^{k-1} [1-(\alpha_W^{SU,1})^2] \|\Delta Z_{m,k-1}^{SU,1}\|^2 \|\widetilde{W}_{m,k}^{SU,1}\|^2$$

$$\leq \|\Pi\| (B_{m,0}^{E,SU,1})^2 + (B_{m,0}^{V,SU,1})^2 - (1-2l_o) \|\Pi\| (B_{m,0}^{E,SU,1})^2 \sum_{i=0}^{k-1} (2k^*)^i$$

$$- \frac{[1-(\alpha_W^{SU,1})^2](B_{m,0}^{V,SU,1})^2}{2} \sum_{i=0}^{k-1} \frac{[1+(\alpha_W^{SU,1})^2]^i}{2^i} \qquad (39)$$

$$\leq \|\Pi\| (B_{m,0}^{E,SU,1})^2 + (B_{m,0}^{V,SU,1})^2 - [1-(2l_o)^k] \|\Pi\| (B_{m,0}^{E,SU,1})^2 - \left[1 - \frac{[1+(\alpha_W^{SU,1})^2]^k}{2}\right] (B_{m,0}^{V,SU,1})^2$$

$$\leq (2l_o)^k \|\Pi\| (B_{m,0}^{E,SU,1})^2 + \left[\frac{1+(\alpha_W^{SU,1})^2}{2}\right]^k (B_{m,0}^{V,SU,1})^2 \qquad \forall k = 1,2,...,N$$

Therefore, the ultimate bounds for SIR error value and value function estimation error can be represented as

$$\|E_{m,k}^{SU,1}\| \leq \sqrt{(2l_o)^k (B_{m,0}^{E,SU,1})^2 + \left[\frac{1+(\alpha_W^{SU,1})^2}{2}\right]^k \frac{(B_{m,0}^{V,SU,1})^2}{\|\Pi\|}} \equiv B_{m.k}^{E,CL} \qquad (40)$$

and

$$\|\Delta Z_{m,k-1}^{SU,1}\|^2 \|\widetilde{W}_{m,k}^{SU,1}\|^2 \leq \sqrt{(2l_o)^k \|\Pi\| (B_{m,0}^{E,SU,1})^2 + \left[\frac{1+(\alpha_W^{SU,1})^2}{2}\right]^k (B_{m,0}^{V,SU,1})^2} \equiv B_{m,k}^{V,CL} \qquad (41)$$

$$\forall k = 1,2,...,N$$

where $B_{m.k}^{E,CL}$ and $B_{m,k}^{V,CL}$ are the values of upper bounds for $k = 1,2,...,N$. According to the representation of (40), (41) and values of $l_o, \alpha_W^{SU,1}$, it is important to note that since tuning parameter $0 < l_o < \frac{1}{2}$ and $0 < \alpha_W^{SU,1} < 1$ given in Theorem 1, then $0 < 2l_o < 1$, $0 < \frac{1+(\alpha_W^{SU,1})^2}{2} < 1$ and terms $(2l_o)^k, \left[\frac{1+(\alpha_W^{SU,1})^2}{2}\right]^k$ will decrease when $k$ increase. Moreover, since initial value function estimation errors $B_{m,0}^{V,SU,1}$ and initial SIR error value $B_{m,0}^{E,SU,1}$ are bounded values, the closed-loop ultimate bounds $B_{m.k}^{E,CL}$ (40) and $B_{m.k}^{V,CL}$ (41) decrease while $k$ increase. Further, when final time $NT_s$ increases, the SIR error values and value function estimation errors will not only be *UUB*, but also these ultimate bounds will decrease with time.

Remark 1: It is important to note since $0 < 2l_o < 1, 0 < \frac{1+(\alpha_W^{SU,1})^2}{2} < 1$ and initial value function estimation error $B_{m,0}^{V,SU,1}$ and SIR error $B_{m,0}^{E,SU,1}$ are bounded values, both term $(2l_o)^k, \left[\frac{1+(\alpha_W^{SU,1})^2}{2}\right]^k$ and the closed-loop ultimate bounds $B_{m,k}^{E,CL}$ (40) and $B_{m,k}^{V,CL}$ (41) will converge to zeros when time goes to infinite (i.e. $B_{m,k}^{E,CL} \to 0$, $B_{m,k}^{V,CL} \to 0$ as $k \to \infty$) and proposed FH-AODPA will achieve infinite optimal DPA solution.

*Case 2: PUs are activated*

Compared with Case 1, closed-loop stability analysis need to be done for SUs and for PUs since the PUs are activated. Similar to Case 1, before introducing the convergence proof, the proposed finite horizon adaptive optimal distributed power allocation algorithm for both PUs and SUs in Case 2 is given as follows.

---

**Algorithm 2:** Finite Horizon Adaptive Optimal Distributed Power Allocation for *lth* PU and *mth* SU in enhanced CRN under Case 2 (i.e. PUs are activated)

---

1: **Initialize:** $\hat{W}_{l,k}^{PU,2} = \mathbf{0}, \hat{W}_{m,k}^{SU,2} = \mathbf{0}$ and implementing admissible policies $\upsilon_{m,0}^{PU,2}, \upsilon_{m,0}^{SU,2}$
2:             for *lth* PU and *mth* SU.
3: **while** $\{kT_s < t < (k+1)T_s\}$ **do**
4:      **Calculate** the value function estimation errors $\Xi_{l,k}^{FTBE,PU,2}, \Xi_{m,k}^{FTBE,SU,2}$ and
5:      $\Xi_{l,k}^{FC,PU,2}, \Xi_{m,k}^{FC,SU,2}$ for *lth* PU and *mth* SU.
6:      **Update** the parameters of the value function estimator for *lth* PU and
7:              *mth* SU as
8:      *lth* PU : $\hat{W}_{l,k+1}^{PU,2} = \Psi_{l,k}^{PU,2}[(\Psi_{l,k}^{PU,2})^T \Psi_{l,k}^{PU,2}]^{-1}[\alpha_W^{PU,2}(\Xi_{l,k}^{FTBE,PU,2})^T$
9:                $+ \alpha_W^{PU,2}(\Xi_{l,k}^{FC,PU,2})^T - (\Gamma_{l,k}^{PU,2})^T - \theta_{m,N}^{PU,2}\Lambda]$
10:    *mth* SU : $\hat{W}_{m,k+1}^{SU,2} = \Psi_{m,k}^{SU,2}[(\Psi_{m,k}^{SU,2})^T \Psi_{m,k}^{SU,2}]^{-1}[\alpha_W^{SU,2}(\Xi_{m,k}^{FTBE,SU,2})^T$
11:              $+ \alpha_W^{SU,2}(\Xi_{m,k}^{FC,SU,2})^T - (\Gamma_{m,k}^{SU,2})^T - \theta_{m,N}^{SU,2}\Lambda]$
12:    **Update** finite horizon adaptive optimal DPA based on estimated $\Theta_{l,k}^{PU,2}$,
13:           $\Theta_{m,k}^{SU,2}$ matrices for *lth* PU and *mth* SU.
14:      *lth* PU : $P_{l,k+1}^{PU,2} = \hat{\upsilon}_{l,k}^{PU,2} I_{l,k}^{PU,2} = -(\hat{\Theta}_{l,k}^{\upsilon\upsilon,PU,2})^{-1}\hat{\Theta}_{l,k}^{\upsilon E,PU,2} z_{l,k}^{PU,2} I_{l,k}^{PU,2}$
15:      *mth* SU : $P_{m,k+1}^{SU,2} = \hat{\upsilon}_{m,k}^{SU,2} I_{m,k}^{SU,2} = -(\hat{\Theta}_{m,k}^{\upsilon\upsilon,SU,2})^{-1}\hat{\Theta}_{m,k}^{\upsilon E,SU,2} z_{m,k}^{SU,2} I_{m,k}^{SU,2}$
16: **end while**
17: **If** $\{t < NT_s\}$ **do**
18:    **Go to** next time interval $[(k+1)T_s, (k+2)T_s)$ (i.e. $k = k+1$), and then go back
19:    to line 2.
20: **else do**
21:    **Stop** algorithm.

---

**Theorem 2.** *(Convergence of the Finite Horizon Adaptive Optimal Distributed Power Allocation for PUs and SUs in enhanced CRN under Case 2).* Let $\upsilon_{l,0}^{PU,2}, \upsilon_{m,0}^{SU,2}$ be any initial admissible policy for *lth* PU's and *mth* SU's finite horizon adaptive optimal DPA scheme in Case 2 with $0 < k^* < 1/2$. Let the *lth* PU's and *mth* SU's parameters be tuned and estimation optimal power allocation be provided by (30) and (32) respectively. Then, there exists positive constant $\alpha_W^{PU,2}, \alpha_W^{SU,2}$ given as $0 < \alpha_W^{PU,2} < 1, 0 < \alpha_W^{SU,2} < 1$ such that the *lth* PU's and *mth* SU's SIR error $e_{l,k}^{PU,2}, e_{m,k}^{SU,2}$ and value function parameter estimation errors $\widetilde{W}_{l,k}^{PU,2}, \widetilde{W}_{m,k}^{SU,2}$ are all *UUB* in Case 2

(i.e. PUs are activated) within the finite time horizon. Moreover, the ultimate bounds are depend on finite time (i.e. $NT_s$), bounded initial value function estimation error $B_{l,0}^{V,PU,2}, B_{m,0}^{V,SU,2}$ and bounded initial SIR error $B_{l,0}^{E,PU,2}, B_{m,0}^{E,SU,2}$.

***Proof:*** Similar to the proofs in Theorem 1.

## 4  Numerical Simulations

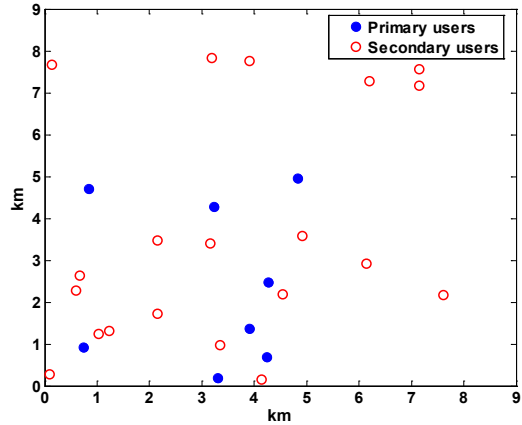

Figure 2. Placement of PUs and SUs in enhanced CRN

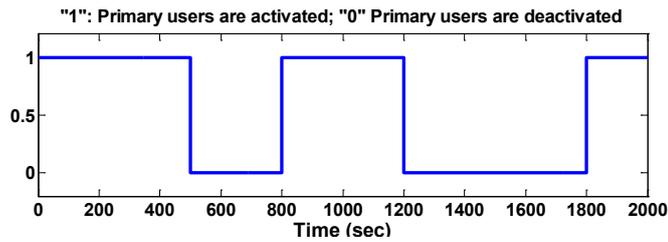

Figure 3. Activity performance of PUs in enhanced CRN

In this simulation, the enhanced Cognitive Radio Network (CRN) is considered to be divided into 2 sub-networks: Primary Radio Network (PRN) and Secondary Radio Network (SRN) which included 8 PUs and 20 SUs. These PUs and SUs are placed randomly within an area of 9km x 9km by using a Gaussian distribution which is shown as Figure 2. Moreover, power of each PU and SU in enhanced CRN is assumed to be updated asynchronously. Therefore, while $l\,th$ PU or $m\,th$ SU updates its power, the powers of all other PUs and SUs do not change. Wireless channel bandwidth $W_c$ is considered to be 100 kHz. Furthermore, it is well known that PUs in enhanced CRN will not always be activated. In this simulation, the activity performance of PUs is shown in Figure 3. According to Figure 3, there exists two cases (i.e. Case 1: PUs are deactivated during $[500\sec,800\sec),[1200\sec,1800\sec)$ ; Case 2: PUs are activated during $[0\sec,500\sec),[800\sec,1200\sec),[1800\sec,2000\sec)$) in enhanced CRN. In Case 1, since PUs are deactivated, a high threshold SIR, $\gamma_H^{SU}$, which each SU tries to achieve is 0.1 (-10dB). In Case 2, to guarantee the QoS of activated PUs, a threshold, $\gamma^{PU}$, which each PU can achieve is selected as 0.1995 (-7dB) and another lower threshold SIR, $\gamma_L^{SU}$, which each SU tries to achieve is set at 0.01 (-20dB).

Next, proposed finite horizon adaptive optimal distributed power allocation (FH-AODPA) is implemented for PUs and SUs in enhanced Cognitive Radio Network with channel uncertainties. Since wireless channel attenuations of users in enhanced CRN are different, initial PUs' and SUs' SIRs are different values (i.e. PUs: [-18.6788dB, -7.8337dB,…, -21.1189dB], SUs: [-8.8116dB, -35.0345dB, -12.4717dB , …, -29.5028dB]). Moreover, the augment SIR error system state is generated as $z_k = [E_k \ \upsilon_k]^T \in \mathbb{R}^{3\times 1}$ and the regression function for value function is generated as

$\{z_1^2, z_1z_2,..., z_2^2,..., z_3^2\}$ as per (19). The design parameter for the value function estimation is selected as $\alpha_W = 0.0001$ while initial parameters for the adaptive estimator are set to zeros at the beginning of simulation.

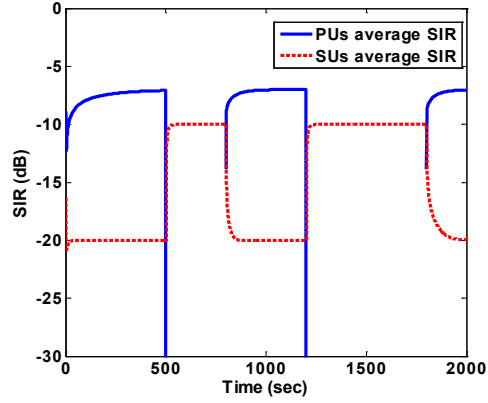

Figure 4. Average SIRs of PUs and SUs in enhanced CRN

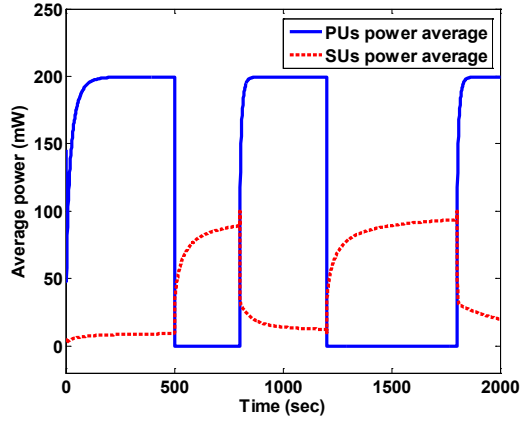

Figure 5. Average power allocation of PUs and SUs in enhanced CRN

In Figures 4 through 9, the performance of proposed finite horizon adaptive optimal distributed power allocation scheme is evaluated. In Figure 4, the averages of all PUs' SIRs and SUs' SIRs are shown. It is important to note that proposed FH-AODPA cannot only force SUs converge to low target SIR (i.e. $\gamma_H^{SU} = -10dB$) when PUs are deactivated (i.e. Case 1: $\gamma_l^{PU} = -\infty \, dB \; \forall l \in \{PUs\}$ during $[500\,\text{sec},800\,\text{sec}), [1200\,\text{sec},1800\,\text{sec}))$, but also force PUs and SUs converge to target SIRs (i.e. $\gamma^{PU} = -7dB, \gamma_L^{SU} = -20dB$) respectively while CRN is at Case 2 (i.e. PUs are activated during $[0\,\text{sec},500\,\text{sec}), [800\,\text{sec},1200\,\text{sec}), [1800\,\text{sec},2000\,\text{sec}))$. Also, the power consumptions averages of PUs and SUs are shown in Figure 5. Obviously, in Case 1, since PUs are deactivated, SUs increase transmission powers to improve network utility (e.g. spectrum efficiency). For Case 2, due to activated PUs, SUs decrease transmission power to reduce the inference to PUs for guarantying their QoS.

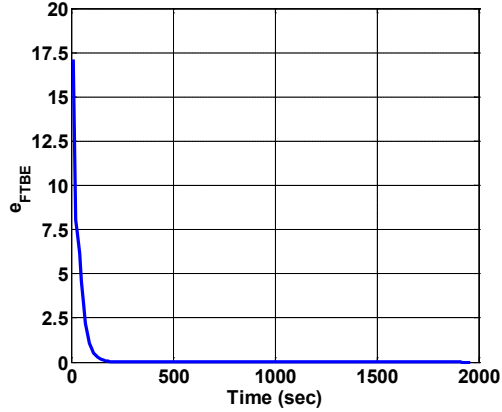

Figure 6. Average of Bellman equation error.

Moreover, the average of Bellman equation errors and terminal constraint errors for both PUs and SUs are considered. As shown in Figure 6 and 7, both average of Bellman equation errors and terminal constraint errors converge close to zeros during the finite horizon (i.e. $t \in [0, NT_s]$ with $NT_s = 2000\,\text{sec}$) which indicates that proposed scheme converges close to optimal power allocation while satisfying the terminal constraint for both PUs and SUs. It is important to note that the convergence performances are dependent upon tuning rate based on Theorem 1 and 2. Further, according to Theorem 1 and 2, when final time $NT_s$ increases, the upper bound of average of Bellman equation errors and terminal constraint errors will decrease.

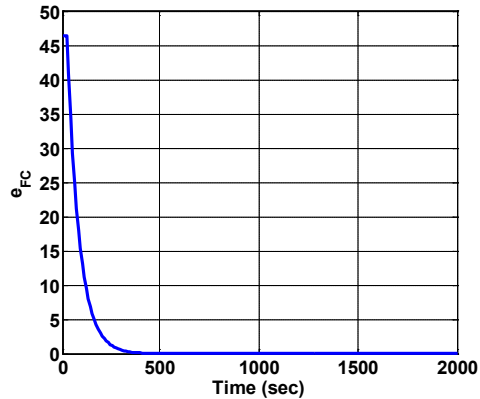

Figure 7. Average of terminal constraint error

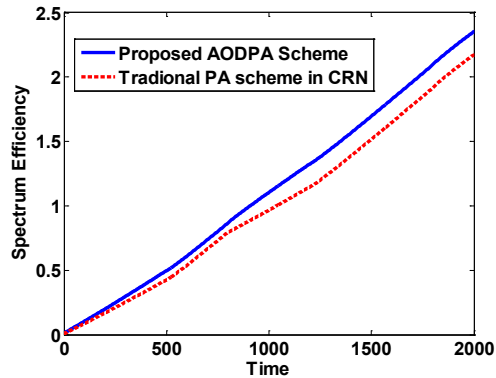

Figure 8. The spectrum efficiency comparison

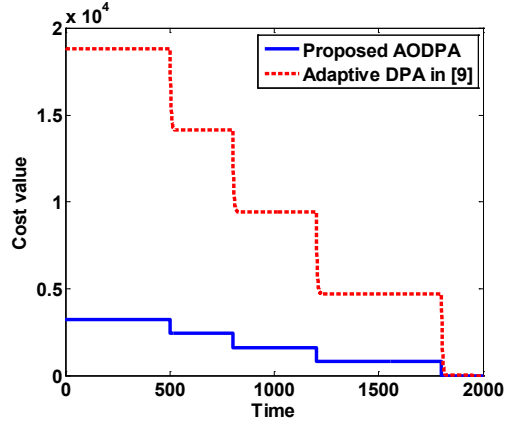

Figure 9. The comparison of cost value

Then, in Figure 8, compared with traditional CRN power allocation schemes [5-6] which prohibit SUs to transmit when PUs are activated, proposed FH-AODPA scheme can increase the spectrum efficiency (i.e. spectrum efficiency (bits/Hz) $= \frac{\text{CRN Throughput (bits)}}{\text{Bandwidth (Hz)}}$ ) by allowing SUs coexist with PUs and allocating power to each user properly. Eventually, for the sake of comparison, adaptive DPA developed in [9] is extended to the enhanced CRN. As shown in Figure 9, proposed FH-AODPA scheme can minimize the cost function (13) more than adaptive DPA in [9]. Therefore, the performance developed FH-AODPA method is better than adaptive DPA [9]. It is important to note that the overshoots always happen at the time two cases switched since target SIRs and SIR errors of PUs and SUs are changed suddenly.

Based on the results presented in Figure 4 through 9, it is important to note the proposed finite horizon adaptive optimal distributed power allocation scheme can not only improve the efficient of enhanced Cognitive Radio Network (e.g. power, spectrum) within finite time horizon, but also does not require the information of channel uncertainties compared with other existing DPA schemes [7-8] in CRN under two cases.

## 5 Conclusion

In this work, the novel SIR error dynamics are developed for both PUs and SUs in enhanced cognitive radio network under two cases (i.e. Case 1: PUs are deactivated, Case 2: PUs are activated) with channel uncertainties. Then, using the SIR error dynamics, a novel finite horizon adaptive dynamic programming scheme is proposed which combines the adaptive estimator (AE) and idea of ADP to solve the Bellman Equation in the real time while satisfying the terminal state constraint and optimize distributed power allocation (DPA) for both PUs and SUs in enhanced CRN under two cases within finite time horizon. The availability of past state values ensured that SIR error dynamics are not needed for proposed FH-AODPA design while an adaptive estimator (AE) generates an estimated value function and a novel finite horizon optimal power allocation law based on the estimation of value function. An initial admissible policy ensures that SIR error systems for PUs and SUs in enhanced CRN are stable for two cases while the adaptive estimator learns the value function and the matrix $\Theta$, and optimal power allocation scheme within finite time horizon. All adaptive estimator (AE) parameters were tuned online using proposed update law and Lyapunov theory demonstrated the *UUB* of the overall closed-loop enhanced CRN system with ultimate bounds which are dependent on final time $NT_s$ and initial system conditions. When the final time is increased, ultimate bounds will be decreased and ultimately converging to zero as time goes to infinite.

**References**

1. Mitola, J., Maguir, G. Q.: Cognitive radio: making software radios more personal. IEEE Personal Communication, vol. 6, pp. 13-18 (1999)
2. Hoang, A. T., Liang, Y. C., Islam, M. H.: Power control and channel allocation in cognitive radio networks with primary users' cooperation. IEEE Transactions on Mobile Computing, vol. 9, pp. 348-360 (2010)
3. Haykin, S.: Cognitive radio: brain-empowered wireless communications. IEEE Journal of Selected Areas on Communication, vol. 23, pp. 201-220 (2005)
4. Kang, X., Liang, Y. C., Arumugam, N., Garg, H., Zhang, R.: Optimal power allocation for fading channels in cognitive radio netwokrs: ergodic capacity and outage capacity. IEEE Transactions on Wireless Communication, vol. 8, pp. 940-950 (2009)
5. Kolodzy, P. J.: Interference temperature: a metric for dynamic spectrum utilization. International Journal of Network Management, vol. 1, pp. 103-113 (2006)
6. Duan, R., Elmusrati, M., Jantti, R., Virrankoski, R.: Power control for time-varying cognitive radio networks. In: Proceeding of $17^{th}$ International Conference on Telecommunication, IEEE Press, pp. 133-137 (2010)
7. Wu, Y., Tsang, H. K.: Distributed power allocation algorithm for spectrum sharing cognitive radio networks with QoS Guarantee, In: Proceeding of $28^{th}$ IEEE International Conference on Computer Communications, IEEE Press, pp. 981-989 (2009)
8. Huang, S., Liu, X., Ding, Z.: Distributed power control for cognitive user access based on primary link control feedback, In: Proceeding of $29^{th}$ IEEE International Conference on Computer Communications, IEEE Press, pp. 1-9 (2010)
9. Jagannathan, S., Zawodniok, M., Shang, Q.: Distributed Power Control of Cellular Networks in the Presence of Rayleigh Fading Channel. In: Proceeding of $23^{rd}$ IEEE International Conference on Computer Communications, IEEE Press, pp. 1055-1066, Hongkong, CHINA (2004)
10. Rappaport, T. S.: Wireless Communication, Principle and Practice. $2^{nd}$ edition. Upper Saddle River, NJ: Prentice Hall (2002)
11. Ren, Z., Wang, G., Chen, Q, Li, H.: Modeling and simulation of Rayleigh fading, path loss, and shadowing fading for wireless mobile networks. Simulation Modeling Practice and Theory, vol. 19, pp. 626-637 (2011)
12. Tse, D., Viswanath, P.: Fundamentals of Wireless Communication, Cambridge University Press, Cambridge, UK (2005)
13. Watkins, C.: Learning From Delayed Rewards. Ph.D. Dissertation, Cambridge University, Cambridge, England (1989)
14. Lewis, F. L., Syrmos, V. L.: Optimal Control. John Wiley and Sons, New York (1999)
15. Jagannathan, S.: Neural Network Control of Nonlinear Discrete-time Systems, CRC Press (2006)
16. Xu, H., Jagannathan, S., Lewis, F. L.: Stochastic optimal control of unknown linear networked control system in the presence of random delays and packet losses. Automatica, vol. 52, pp. 1017-1030 (2012)
17. Brochett, R. W., Millman, R.S., Sussmann, H. J.: Differential Geometric Control Theory, Birkhauser, USA (1983)